\documentclass[graybox]{svmult}

\usepackage{type1cm}        
\usepackage{makeidx}         
\usepackage{graphicx}        
\usepackage{multicol}        
\usepackage[bottom]{footmisc}

\usepackage{newtxtext}       %
\usepackage{newtxmath}       

\usepackage{booktabs}
\usepackage[english]{babel}
\usepackage{url}
\usepackage{xcolor}
\usepackage[export]{adjustbox}
\usepackage{colortbl}
\makeatletter
\def\hlinewd#1{%
  \noalign{\ifnum0=`}\fi\hrule \@height #1 \futurelet
   \reserved@a\@xhline}
\makeatother

\UseRawInputEncoding
\usepackage{subcaption}

\usepackage{graphicx}

\usepackage[T1]{fontenc}

\usepackage{hyperref}
\hypersetup{
    bookmarks=false,         
    unicode=false,          
    pdftoolbar=true,        
    pdfmenubar=true,        
    pdffitwindow=false,     
    pdfstartview={FitH},    
    pdftitle={ECAI 2016},    
    pdfauthor={},     
    pdfsubject={},   
    pdfcreator={},   
    pdfproducer={}, 
    pdfkeywords={}, 
    pdfnewwindow=true,      
    colorlinks=true,       
    linkcolor=blue!80!black,          
    citecolor=blue!80!black,        
    urlcolor=blue!80!black,           
    filecolor=black      
}

\usepackage[margin=1in]{geometry}

\usepackage{pbox}
\usepackage{natbib}

\makeindex             

\begin{document}

\title*{{\sffamily{Visuo-Locomotive Complexity as a Component of\\Parametric Systems for Architecture Design}}}

\author{Vasiliki Kondyli, $~$Mehul Bhatt $~~$(\"{O}rebro University, Sweden)\\Evgenia Spyridonos $~~$(University of Stuttgart, Germany)}



\authorrunning{$~$} 

\institute{\textbf{CoDesign Lab}$~~~$/$~~~$\url{www.codesign-lab.org}$~~$>$~~$\url{info@codesign-lab.org} $~~~~$---$~~~~$ The DesignSpace Group$~~$/$~~$\url{www.design-space.org}}


\maketitle

\abstract{$~~${\normalsize\sffamily A people-centred approach for designing large-scale built-up spaces necessitates systematic anticipation of user's embodied visuo-locomotive experience from the viewpoint of human-environment interaction factors pertaining to aspects such as navigation, wayfinding, usability. In this context, we develop a behaviour-based  visuo-locomotive complexity model that functions as a key correlate of cognitive performance vis-a-vis internal navigation in built-up spaces. We also demonstrate the model's implementation and application as a parametric tool for the identification and manipulation of the architectural morphology along a navigation path as per the parameters of the proposed visuospatial complexity model. We present examples based on an empirical study in two healthcare buildings, and showcase the manner in which a dynamic and interactive parametric (complexity) model can promote behaviour-based decision-making throughout the design process to maintain desired levels of visuospatial complexity as part of a navigation or wayfinding experience.}
}
\medskip
\medskip


\vfill

{\large\sffamily\textbf{Publication Note}}
\medskip
{\sffamily 

This is a preprint of the contribution published as part of the proceedings of ICoRD 2021: 8th International Conference on Research into Design, IDC School of Design (IIT Mumbai, India).\\ICoRD 2021, \url{www.idc.iitb.ac.in/icord2021/}.

\medskip
\medskip

The overall scientific agenda driving this research may be consulted here:\\The DesignSpace Group$~~$/$~~$\url{www.design-space.org}

\medskip
\medskip

Select publications related to this research are available at:\\\url{www.codesign-lab.org/select-papers} (under DesignSpace)
}

\newpage

\section{Introduction}
The design of navigation and wayfinding systems within large-scale built-up spaces is a particularly challenging task: research in behavioural, computational, and mixed-methods interdisciplinary approaches in spatial cognition, architecture design cognition, and artificial intelligence for design have shown that diverse morphological and ecological aspects pertaining to space, perception, and human factors constitute a correlate of visuo-locomotive cognitive experience in built-up space from the viewpoint of tasks as navigation and wayfinding \citep{BhattChapter2017,Bhatt-ICSC-2018,DBLP:conf/kr/BhattST14,Kondyli2018Rotation,Devlin2014,ONeill1991,Weisman1981}.

\medskip

\begin{figure}[t]
\centering
\includegraphics[width=1\textwidth]{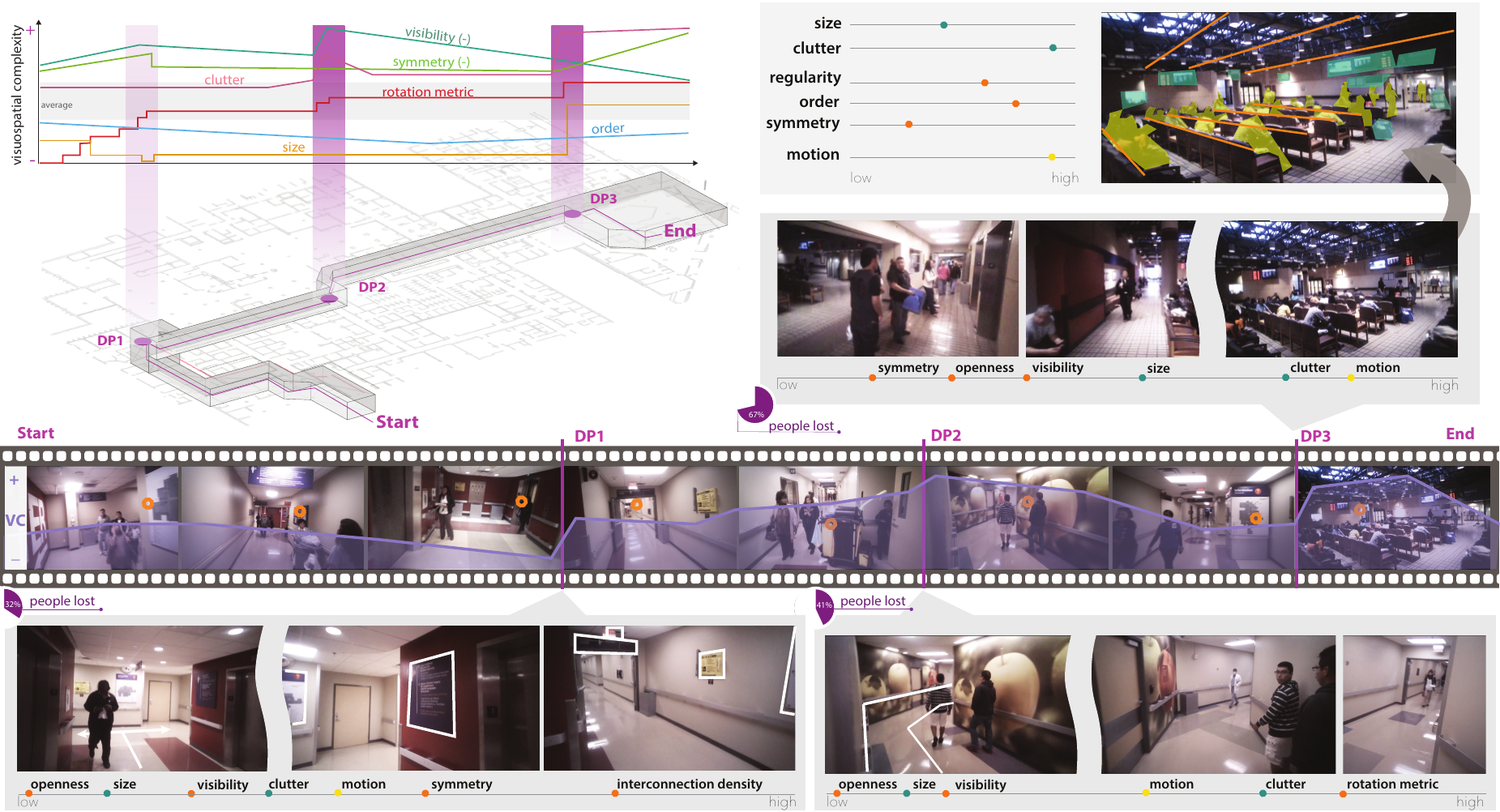}
\caption{\small{\textbf{Large-Scale Evidence Based Analysis of Visuo-Locomotive Experience}.\quad Analysis of visuospatial complexity as it evolves along the navigation path at the Old Parkland Hospital. Analysis of decision points (DP) where participants get lost reveal the combination of attributes that compose the current visuospatial complexity level (colors correspond to the categories of attributes of Table \ref{table}: Quantitative (green), Structural (orange), Dynamic (yellow)).}}
\label{intro}
\end{figure}

 \textbf{Visuo-Locomotive Experience During Navigation} \quad Visuo-locomotive experience most broadly refers to the embodied cognitive experience of environmental space through a multimodal perceptual integration, e.g., encompassing visual, locomotive and auditory exploration  \citep{BhattChapter2017,DBLP:conf/kr/BhattST14}.  From the viewpoint of this paper, examining the embodied experience of users through interaction with the architectural space during active locomotion includes quantitative measurements and qualitative analysis of a range of aspects including visual attention (e.g. fixation, saccades),  fine-grained behaviour (e.g., stopping, looking around, interacting with other people), orientation tasks, spatial memory (e.g. sketch maps, questionnaires) etc  (details in papers \citet{DBLP:conf/aaai/BhattSSKG16,Kondyli2018Rotation}). The focus of this paper is on visuo-locomotive experience as it pertains to aspects of visuospatial and locomotive complexity as articulated in this research.
 \medskip 
  
  \textbf{Visuospatial and Locomotive Complexity} \quad   To measure the effect of the architectural space on the visuo-locomotive experience we take into consideration  behavioural data of perceived visuospatial stimuli with respect to the attributes involved in the dynamic scene (e.g. form, colours, layout) (Fig. \ref{intro}). Even though a combination of these attributes can result to complex scenes, people are able to form a coherent percept amid numerous regions and identify properties and semantics of a scene at a glance \citep{Potter1976}. Visual complexity has been  broadly defined as the level of detail and intricacy contained within the image or a scene \citep{Snodgrass1980}. Its effect on visual attention and spatial cognition, and corresponding measurement methodologies have been investigated in disciplines such as cognitive science, psychology, computer science, or marketing \citep{Cavalcante2014,Moacdieh2015,Rosenholtz2007,Madan2018}.  With the term \emph{visuospatial and locomotive complexity} we consider the combination of visual and spatial characteristics that both coexist in dynamic naturalistic scenes where a person navigates \citep{Kondyli2018Rotation,KondyliSTAIRS20} \footnote{Henceforth, we use the terms visuospatial and visuo-locomotive complexity interchangeably. Other factors influencing embodied visuo-locomotive experience in space are familiarity, nature of the task (e.g. exploration, wayfinding), individual differences (e.g. age, spatial skills) etc. However these aspects are beyond the scope of this paper.}. Although complexity can provide an interesting and cognitively stimulating environment  \citep{Cassarino2016}, excessive visual or spatial complexity can impact cognitive processes involved in visuo-locomotive experience of people in everyday tasks such as navigation or wayfinding  \citep{Boeing2018,Fortenbaugh2008,Feldman2004}.  

\medskip

In this paper we study the visuospatial complexity of the architectural space from the human-centred perspective of embodied visuo-locomotive experiences. We present a  cognitive model of visuospatial and locomotive complexity that embeds empirical knowledge on the effect of environmental and dynamic attributes on human behaviour and we define a scale of complexity. We introduce the model to a parametric design system and demonstrate how a designer can identify \emph{Classets} of complexity for a particular path or a segment of it, as well as modify the morphology of space along a path in order to maintain complexity to a preferable level.  As currently there is no universally accepted or empirically defined scale for visuospatial complexity our method aims to organise and link the different attributes involved and their contribution to a visuospatial complexity scale, however, more empirical studies are needed to support our classification.

\section{A Model of Visuo-Locomotive Complexity for Spatial Design}

To model visuospatial complexity based on its effect on humans during active locomotion, we examine the attributes involved in the naturalistic dynamic scene and how they evolve along a navigation path. We combine empirical results from previous studies on visual and spatial complexity metrics together with the result of a behavioural study in two healthcare facilities in the Parkland Hospital \citep{KondyliEGICE2018}. We consider visuospatial and locomotive complexity as an implicit measure of cognitive load \citep{Harper2009} and inefficient navigation, and we use the behavioural analysis along the path (through a combination of behavioural metrics such as delays, visual search performance, ask for help, loss of orientation, etc.) to identify locations with high complexity and further examine the present environmental and dynamic attributes. 

\medskip
\medskip
\medskip

{
\begin{table}[t]
\renewcommand{\arraystretch}{1.2}
\begin{center}
 \scriptsize\sffamily
\begin{tabular}{>{\columncolor[gray]{0.92}} p{4.5cm} p{9cm}}
\hlinewd{1pt}
\rowcolor[gray]{.92}\textbf{{\color{blue!90!black}VISUOSPATIAL COMPLEXITY}} & \textbf{Proportional relationship to complexity and  Description}  \\\hline\hline
\multicolumn{2} {l} {\textbf{{\color{blue!90!black}A1}. \quad Quantitative Attributes}} \\\hline

 SIZE & (+/-) \quad  The dimensions of the physical space, the area coved by the visual stimulus.  \\[6pt]
 
 CLUTTER   & (+) \quad   \\
 Quantity   &   (+) \quad No. components  (objects, people, shapes etc.) \\
 Variety of Colours &  (+) \quad No. colours \\
 Variety of Shapes/Objects & (+) \quad No.  shapes / objects   \\ 
 Objects Density & (+) \quad No. objects in a defined area \\
 Edges Density &  (+) \quad No. edges of objects, No. of polygon vertices of a perimeter (fractal dimension)  \\
 Luminance & (-) \quad  Amount of light emitted / reflected from the scene \\
Saliency & (-) \quad Prominent elements based on characteristics of colour, luminance and contrast   \\ 
Target-background similarity & (+) \quad Compare similarity in luminance, contrast, structure, orientation, etc. \\

\hline
\multicolumn{2} {l} {\textbf{{\color{blue!90!black}A2}. \quad Structural  Attributes}} \\\hline

Repetition & (-) \quad Recurrence of (groups) elements or characteristics on a line/a grid/ a pattern \\
Symmetry & (-) \quad  Resilience to transformation (reflectional/rotational/translational/helical/fractal)   \\ 
Order & (-) \quad Organised elements based on a recognised structure, fractal dimension, axial  lines \\
Homogeneity/Heterogeneity  & (-) \quad Being all the same kind or diverse (single shape repeated - multiple distinct shapes) \\
Regularity  & (-) \quad Variations in a placement rule across a surface/line (polygons - abstract shapes)\\
Openness &  (-) \quad The ratio between empty and full space  \\ 
Grouping &  (-) \quad No. elements that are part of a group  \\
Rotation Metric$^{*}$   & (+) \quad Accumulated degrees of rotation angle during locomotion, No. of turns \\
Visibility$^{*}$ & (-) \quad  Visual range from a vantage point, visual connectivity between points\\
Interconnection Density$^{*}$ & (+) \quad  No. of directional choices in each node (e.g. decision point, junction)\\
\hline
\multicolumn{2} {l} {\textbf{{\color{blue!90!black}A3}. \quad Dynamic  Attributes}} \\\hline

Motion &  (+) \quad No. people or objects moving in the scene \\ 
Flicker   &    (+) \quad  Abrupt changes over-time (in luminance, colours, etc.)  \\
Speed | Direction &  (+) \quad  The rate of change of position with respect to time | Move or facing towards \\

\hlinewd{1pt}
\end{tabular}
\caption{\small{ \textbf{Taxonomy of Attributes for Visuospatial Complexity}.\quad The relationship between the individual attributes and the visuospatial complexity is defined as proportional (+) or inversely proportional (-) according to the impact they have on the visuospatial complexity scale. The  ($^{*}$) indicates locomotive complexity attributes.}}
\label{table}
\end{center}
\end{table}
}

\textbf{Measuring Visuo-Locomotive Experience}\quad  The model (Table \ref{table}) derives from a  taxonomy of objective physical attributes from the scene analysis that affects visuospatial perception and cognitive functions (e.g. driving, walking)  \citep{KondyliSTAIRS20}. However, embodied cognitive experience whilst in motion along a path cannot be investigated based on a discrete analysis of a series of singular scenes. Therefore, the model is enriched with attributes pertaining to visuo-locomotive complexity such as rotation metric, visibility, interconnection density \citep{Kondyli2018Rotation}. The attributes are categorised into:
\medskip

{\color{blue!90!black}\textbf{A1}.}$~~$\emph{Quantitative Attributes.}\quad Referring to objective environmental attributes such as low-level (edge, colors) and middle-level (corners, orientation) features of the scene. Studies on visual attention mostly on real-world static scenes, have shown how these environmental attributes can lead to a overabundance of information (known as clutter), work as distractors, and impact the visual search performance \citep{Henderson2009, Rosenholtz2007,Park2014}. Clutter in architectural space may also involve the fractal dimension, referring to the number of polygon vertices of a perimeter of a space. Manifest cues, such as signage and landmarks, are part of clutter however because of their semantic significance as navigation aid objects they constitute major visual attention targets. 
\medskip

{\color{blue!90!black}\textbf{A2}.}$~~$\emph{Structural Attributes.}\quad Refer to the relation that the elements form due to positioning in space, or the overall distribution on the viewing scene. The presence of regularities, symmetry, repetition, or order simplifies the scene \citep{Feldman2004, vanderHelm2000} while randomised arrangements contribute to higher complexity classes. Homogeneity/Heterogeneity, regularity of shapes and objects, grouping of elements in space, openness of space and their relationship to complexity have been previously introduced in the fields of urban design and architecture \citep{Boeing2018, Salingaros2014}. These attributes formulate the environmental structure along the navigation path, and impact the holistic legibility of the environment, including aspects of interconnection density  \citep{Weisman1981,ONeill1991}, visibility (e.g. visual range, line of sight) \citep{Benedikt1979}, rotational locomotion \citep{Kondyli2018Rotation}. On the contrary, wayfinding studies show the value of architectural differentiations and breaks in the structural order as a design tool for distinguished areas that facilitate navigation \citep{Baskaya2004}. 

 \medskip

{\color{blue!90!black}\textbf{A3}.}$~~$\emph{Dynamic Attributes.}\quad Studying active embodied locomotion, such as navigation, involves studying dynamic parts of the scene, as for example people or moving objects or the ratio of changes per meter. Dynamic attributes have a major impact on visual attention patterns, as clusters of attention often coincide with semantically rich objects such as eyes, hands \citep{Mital2011}. Moreover, cortical analysis shows a selective response to moving elements on the scene, meaning that people are able to notice moving objects even if they are not looking for them \citep{Rosenholtz2007}.

\begin{figure}[t]
\begin{center}
\includegraphics[width=1\textwidth]{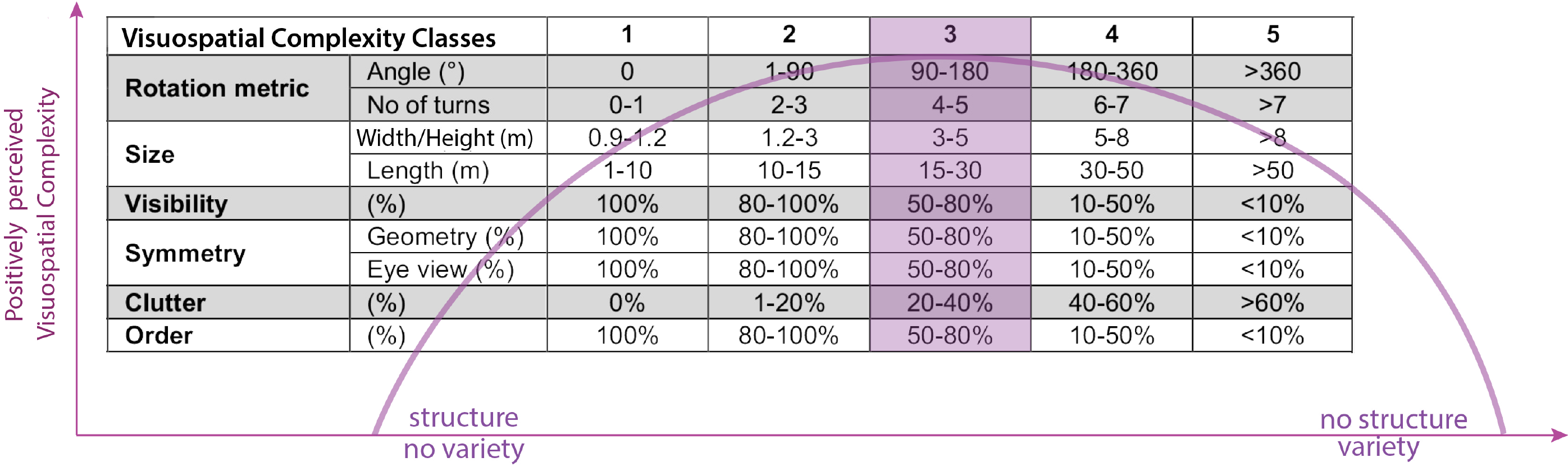}
\caption{\footnotesize \textbf{Definition of Visuospatial Complexity Classes.} Simplified version of Visuospatial Complexity Classes developed on selected attributes for the purpose of parametric modelling (Section \ref{parametric}). The curve represents an interpretation of the overall positively  perceived visuospatial complexity as a function of variety and  structure that picks on a moderate Class {3}  (adapted from Shiner  et. al. \citep{Shiner1999}).} 
\label{table_Classes}
\end{center}
\end{figure}

\section{An Empirical Basis for a Visuospatial Complexity Scale} 

\textbf{Case study in Parkland Hospital}\quad In a behavioural case study conducted in two large-scale healthcare facilities (Old and New) in Parkland Hospital in Texas (USA), 25 participants performed a navigation task from the Emergency room to the Pharmacy. We employed a range of sensors for measuring embodied visuo-locomotive experience and navigation performance (e.g. eye-tracking, egocentric video, external camera based video, questionnaires, orientation task) and detect locations in the path where a number of participants were disoriented or confused (details in  \citep{DBLP:conf/aaai/BhattSSKG16,KondyliEGICE2018}). 

 \medskip

\textbf{Behavioural and Environmental Analysis}\quad The correlations between the environmental and the behavioural analysis suggest that peculiar geometries (non symmetrical, not regular), narrow, long and high cluttered corridors are places where people tend to be frequently confused. Additionally, the presence of a combination of these attributes leads to low visibility and a great angle of rotation by the participant along the path that aggravate the navigation performance. For instance, in Decision Point 2 in the Old Parkland Hospital, we record 41\% of participants having a disorientation or confusion event (e.g.  stop,  hesitation,  look  around,  intensive  visual search to both directions, etc.  \citep{Kondyli2018Rotation}, Fig. \ref{intro}). The place is characterised by narrow non-symmetrical and non-regular long cluttered corridors, that leads to reduced visibility. The place is also regularly occupied by dynamic attributes such as moving obstacles, pedestrian, people on wheelchairs moving in various directions. However, the questionnaires show that the landmark in the corridor (apple posters in bottom row of Fig. \ref{intro}) was visually accessible and well recorded in spatial memory.
 
\medskip

\textbf{The effect of combined attributes}\quad These observations suggest that the combination of attributes present in the scene can have a counterbalancing effect on the overall visuospatial complexity level, however further empirical work is needed to define the combinations of attributes that can lead to this outcome. For instance, high clutter increases the visuospatial complexity level but it can be mitigated with a well organised structural scene. Another consideration about finding balance in the visuospatial complexity scale is inspired by convex function of disorder \citep{Shiner1999} and complex adaptive systems \citep{Gershenson2012}, suggesting that a positive effect from the perceived visuospatial complexity is expected to reach an optimum between variety and structure  (Fig. \ref{table_Classes}).  For example, highly symmetrical spaces with repetitive features are related to low visuospatial complexity, but they have a negative effect on navigation performance as there is no architectural differentiation or distinctive object to assist cognitive mapping \citep{Baskaya2004}. Consequently, in the process of developing a scale for visuospatial complexity, we consider that humans prefer to experience information at a comfortable rate (too little deprives the senses and too much overloads them \citep{Boeing2018}), and so we suggest a moderate visuospatial complexity class as a design aim (Fig. \ref{table_Classes}). This design aim should be adjusted based on the specific design groups (e.g. children, older adults, people with cognitive decline) in order to provide a suitable motivating environment.

\medskip

\textbf{Empirically model complexity Classes}\quad As an example of the modelling process we focus on selected attributes for demonstration reasons, and we define in approximation the Classes based on the behavioural analysis of the navigation study \citep{Kondyli2018Rotation,KondyliEGICE2018}. The Classes are defined based on the empirical results and the proportional relation between each attribute and visuospatial complexity (Table \ref{table}). Further empirical work is needed for the verification of the numeric and percentage classification (Fig. \ref{table_Classes}). Specifically, \emph{Rotation metric} is defined by the number of turns along the path (or along a segment of the path) as well as by the accumulating degrees of the angle the person performs while navigating. For the definition of \emph{Size}, we combine the three dimensions of space and calculate the average along the path or the segment. \emph{Visibility} is defined based on the percentage of the visual range a navigator has while walking from the different segments towards the end of the path. \emph{Symmetry} is calculated in respect to the layout geometry, as well as by analysing the symmetry in the 3D scene the navigator encounters along the path. \emph{Clutter} is defined by the percentage of space covered by 3D elements in the total path (or each segment), and  \emph{Order} refers to the related organisation of these elements in space, calculated by the percentage of the area that reflects a known structure (e.g. grid, circle, rectangular, spiral, square). To calculate the visuospatial complexity class in an aggregative level for the overall path, we combine the corresponding values of each attribute and provide an average result with a numerical indication between {1} and {5} (Fig. \ref{table_Classes}).

\begin{figure}[t]
\begin{center}
\includegraphics[width=1\textwidth]{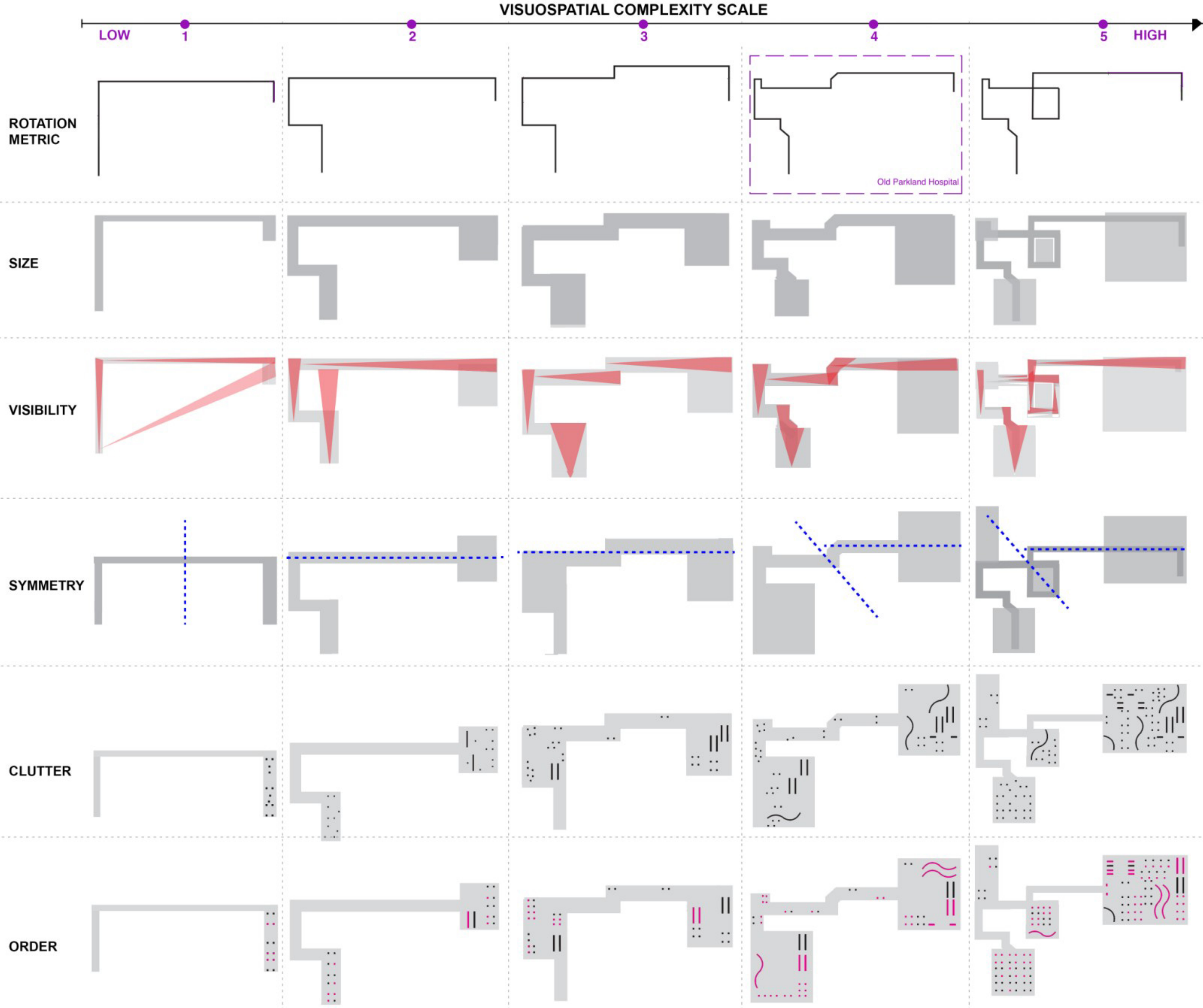} 
\caption{\small \textbf{Visuospatial Complexity Scale}.\quad Manipulation of the path in Old Parkland Hospital based on the scale of visuospatial complexity (Fig. \ref{table_Classes}) and the selected attributes that change the morphology respectively. The purple dashed rectangular is the starting point of the identification function of the given path geometry.}
\label{identification}
\end{center}
\end{figure}

\section{Parametric Modelling of Visuospatial Complexity} \label{parametric}

By parametrising selective attributes in respect to the scale of visuospatial complexity, and monitoring the relations between them, we develop a parametric model that runs two main operations:  \emph{identification} of the visuospatial complexity class of a given morphology, and  \emph{manipulation} of the morphology to correspond to a preferred class (e.g. moderate Class {3}, Fig. \ref{table_Classes}). We approach the parametric model from the perspective of the human-navigator, using the navigation path as the first component and the main geometrical input to the model. The parametric tool promotes interactive design decision-making based on the visuospatial complexity model (Table \ref{table}) that can be useful for: comparing different versions of a design in terms of the visuospatial complexity Classes, identifying the class  and manipulating the morphology in relation to a selected attribute or in aggregative level for all attributes, comparing buildings in relation to their visuospatial complexity Class, manipulating morphologies to eliminate or increase the visuospatial complexity Class. 

\medskip

 \begin{figure}[t]
\begin{center}
\includegraphics[width=1\textwidth]{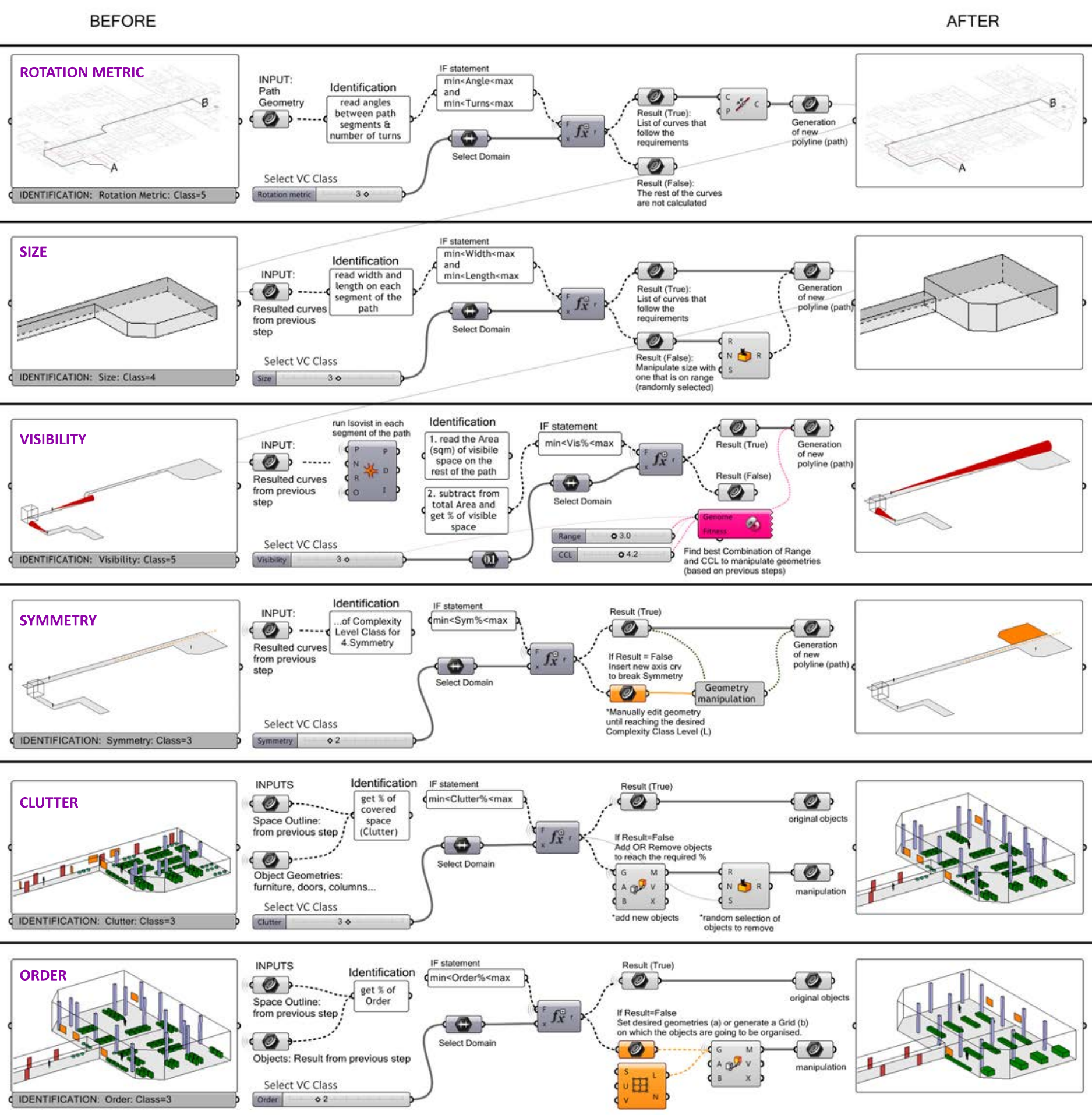}
\caption{\footnotesize \textbf{Parametric Process for Manipulation of Visuospatial Complexity Class}.\quad An example of the manipulation function for the path at the Old Parkland Hospital (Fig. \ref{intro}),  implemented in Rhino CAD-Grasshopper3D. On the left the result of the identification function reporting the visuospatial complexity  (VC) class per attribute, and on the right the result of the manipulation function in steps, monitoring the morphological changes and aiming at an average moderate complexity class for the overall path.}
\label{manipulation}
\end{center}
\end{figure}

\textbf{Class Identification and Morphology Manipulation} \quad  The identification function receives  as an input a polyline representing the navigation path, it analyses the characteristics of the path and categorises them into Classes based on the visuospatial complexity model (e.g. purple dashed rectagular in Fig. \ref{identification}). By manipulating the morphology of space along the navigation path, we can expand or eliminate the visuospatial complexity by ``controlling'' the parametric ``sliders'' between the defined Classes of each attribute (Fig. \ref{identification}). In the Old Parkland Hospital example (Fig. \ref{manipulation}), we aim at maintaining a moderate Class {3} of visuospatial complexity along the path. The identification function gave the result of an overall visuospatial complexity Class {4}. The manipulation function (based on the selected group of attributes) can be used to introduce changes in the morphology (e.g. geometry, arrangement of elements), to reduce the overall visuospatial complexity. Specifically, \emph{Rotation metric} simplifies the geometry of the path based on the angles of turns between the segments.  The \emph{Size} function modifies the three dimensions of the segments of the path to fit the requirements of the Class. The \emph{Visibility} reports the percentage of visible space from each segment towards the end point of the path and suggests alternative combination between visual range and Classes using an evolutionary solver (Galapagos for Grasshopper3D). The \emph{Symmetry} function introduces new symmetry axes to manipulate the geometry in order to fit the requirements of the respective Class, while the \emph{Clutter} and \emph{Order} functions, add or remove objects and organise them in groups.

\begin{figure}[t]
\begin{center}
\includegraphics[width=1\textwidth]{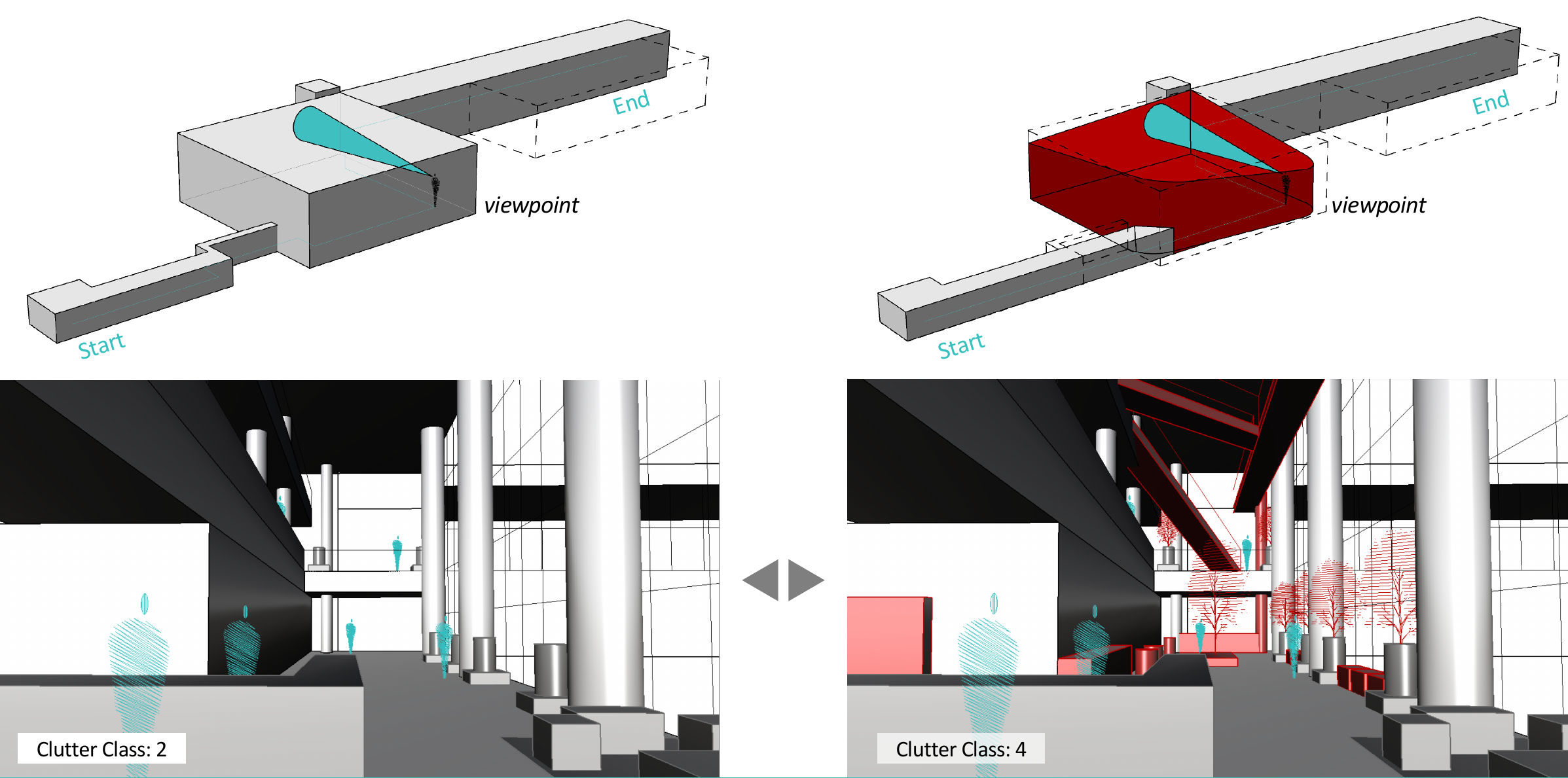} 
\caption{\small \textbf{Manipulation of the Morphology of the LLobby at the New Parkland Hospital.}\quad We modify the visuospatial complexity for the specific segment of the path for the attribute of \emph{Clutter}, while we make sure that the rate visuospatial complexity for the overall path remains at Class 3.}
\label{NPH}
\end{center}
\end{figure}

The manipulation function can be used in the overall or in particular segments of the path to moderate, or eliminate the visuospatial complexity level. However, it can also contribute to indicating areas where visuospatial complexity can be increased to fulfil design requirements while the designer can closely monitor how these changes affect the visuospatial complexity level. For instance, the analysis of the path at the New Parkland Hospital suggests low overall visuospatial complexity (Class {2}). We use the manipulation functionality for the segment of the path that involves the entrance lobby and we extent the level of visuospatial complexity from Class 2 to Class 4 without interrupting the average complexity level for the overall path. Thus, we can reassure an acceptable navigability experience for the users while experimenting with objects at the entrance lobby  (Fig. \ref{NPH}).

\section{Summary}

Architects are confronted with anticipating the effect of their design decisions on users' cognitive experience ---termed \emph{visuo-locomotive experience}--- as they move along a navigation path. Towards this end,  cognitive assistive technologies for design can be useful during the different steps of the design process if they incorporate empirical knowledge from behavioural studies in real-world built environments \citep{BhattChapter2017,DBLP:conf/kr/BhattST14,KondyliEGICE2018,Kondyli2018Rotation}. Parametric design systems provide the flexibility and adaptability needed, however currently their established functions are numerically and geometrically oriented. By introducing the dimension of human behaviour in parametric design systems, such as high-level cognitive design requirements, and parameters of morphological formulation emanating therefrom, we can promote people-centred architecture design from the first steps of the design process \citep{BhattChapter2017,KondyliEGICE2018}.
\medskip

The ability to analyse and manipulate  the morphology of an environment (and resulting navigation paths) based on people-centred aspects, such as the scale of visuospatial and locomotive complexity and its effects on visuo-locomotive experience, is useful for the evaluation of  design decisions and prediction of design performance especially for the initial stages of the design process. We have demonstrated how a parametric design system that incorporates behavioural knowledge can facilitate early design decisions especially for large-scale built environments such as hospitals, airports and museums where it is crucial to predict navigation performance of users at varying levels of visuo-locomotive complexity.

\medskip
Identification of suitable attributes to quantify an optimal level of visuospatial complexity for better cognitive performance (e.g., during navigation) needs to be investigated by empirical work. Towards this direction, our next steps include the use of the parametric functions developed to generate a series of immersive virtual environments that systematically differ in the class of visuospatial complexity the proposed model. In a series of case studies in VR we employ behavioural metrics  (e.g. behavioural mapping, eye-tracking, orientation task, etc.) to measure the effect of the defined attributes and their combinations on the navigation performance and better inform our model of visuo-locomotive  complexity.

\newpage

\nocite{DBLP:conf/apgv/BhattSKS16,DBLP:conf/aaai/BhattSSKG16,10.1007/978-94-017-9297-4_7,DBLP:conf/vl/BhattSH12,DBLP:conf/cosit/BhattDH09}
\bibliographystyle{abbrvnat}


\end{document}